\documentclass[twocolumn,secnumarabic,amssymb, nobibnotes, aps, prl,superscriptaddress]{revtex4-1}
\usepackage[colorlinks=true,linkcolor=blue,citecolor=blue,urlcolor=blue]{hyperref}
\usepackage[separate-uncertainty=true]{siunitx}
\usepackage{graphicx}
\usepackage{gensymb}
\usepackage{amsmath,bm}
\usepackage{braket}
\renewcommand{\Re}{\operatorname{Re}}
\renewcommand{\Im}{\operatorname{Im}}

\newcommand{\angstrom}{\mbox{\normalfont\AA}}

\begin{document}

\title{Attosecond time-domain measurement of core-excitonic decay in magnesium oxide}%
\author{Romain G\'{e}neaux}%
\thanks{These two authors contributed equally}
\email{rgeneaux@berkeley.edu}
\affiliation{Department of Chemistry, University of California, Berkeley, 94720, USA}
\author{Christopher J. Kaplan,$^*$}
%\thanks{These authors contributed equally to this work}
\affiliation{Department of Chemistry, University of California, Berkeley, 94720, USA}

%\altaffiliation{These authors contributed equally to this work}
%
\author{Lun Yue}
\affiliation{Department of Physics and Astronomy, Louisiana State University, Baton Rouge, Louisiana 70803, USA}

\author{Andrew D. Ross}
\affiliation{Department of Chemistry, University of California, Berkeley, 94720, USA}

\author{Jens E. B{\ae}kh{\o}j}
\affiliation{Department of Physics and Astronomy, Louisiana State University, Baton Rouge, Louisiana 70803, USA}

\author{Peter M. Kraus}
\affiliation{Department of Chemistry, University of California, Berkeley, 94720, USA}

\author{Hung-Tzu Chang}
\affiliation{Department of Chemistry, University of California, Berkeley, 94720, USA}
\author{Alexander Guggenmos}
\affiliation{Department of Chemistry, University of California, Berkeley, 94720, USA}
\author{Mi-Ying Huang}
\affiliation{Department of Chemistry, University of California, Berkeley, 94720, USA}
\author{Michael Z\"{u}rch}
\affiliation{Department of Chemistry, University of California, Berkeley, 94720, USA}
\author{Kenneth J. Schafer}
\affiliation{Department of Physics and Astronomy, Louisiana State University, Baton Rouge, Louisiana 70803, USA}

\author{Daniel M. Neumark}
\affiliation{Department of Chemistry, University of California, Berkeley, 94720, USA}
\affiliation{Chemical Sciences Division, Lawrence Berkeley National Laboratory, Berkeley, California 94720, USA}

\author{Mette B. Gaarde}
\affiliation{Department of Physics and Astronomy, Louisiana State University, Baton Rouge, Louisiana 70803, USA}

\author{Stephen R. Leone}
\email{srl@berkeley.edu}
\affiliation{Department of Chemistry, University of California, Berkeley, 94720, USA}
\affiliation{Department of Physics, University of California, Berkeley, 94720, USA}
\affiliation{Chemical Sciences Division, Lawrence Berkeley National Laboratory, Berkeley, California 94720, USA}

\begin{abstract}
Excitation of ionic solids with extreme ultraviolet pulses creates localized core-excitons, which in some cases couple strongly to the lattice. Here, core-excitonic states of magnesium oxide are studied in the time domain at the Mg $\text{L}_{2,3}$ edge with attosecond transient reflectivity spectroscopy. Attosecond pulses trigger the excitation of these short-lived quasiparticles, whose decay is perturbed by time-delayed near infrared optical pulses. Combined with a few-state theoretical model, this reveals that the optical pulse shifts the energy of bright core-exciton states as well as induces features arising from dark core-excitons. We report coherence lifetimes for the first two core-excitons of $2.3 \pm 0.2$ and $1.6 \pm 0.5$ femtoseconds and show that these short lifetimes are primarily a consequence of strong exciton-phonon coupling, disclosing the drastic influence of structural effects in this ultrafast relaxation process.
\end{abstract}

\maketitle
Excitation of a solid with a high-energy photon leads to an ultrafast dynamic response involving structural and electronic degrees of freedom. A common spectroscopic approach to study these processes is to compare the absorption and emission spectra of the system, which respectively reveal which states were excited and eventually populated, as is done for instance in resonant inelastic x-ray scattering (RIXS) \cite{Ament2011}. This grants indirect access to the behavior of the transient core-excited state.
In contrast, the development of attosecond science now offers an unmediated view of relaxation processes \cite{Krausz2009a,Ramasesha2016a}. By measuring the system in the time domain, the intermediate states involved in the decay can be directly probed. For instance in attosecond transient absorption spectroscopy \cite{Kraus2018,Geneaux2019}, short extreme ultraviolet (XUV) pulses can be used to precisely trigger the core-hole excitation, whose decay is then tracked with optical femtosecond laser pulses \cite{Chew2018,Beck2014}. %Attosecond techniques are therefore complementary time-domain methods to spectral measurements such as RIXS.

In the condensed phase, an important question arises: can nuclear motion influence the decay of the core-hole for states living only a few femtoseconds? Intuitively, one might expect that if the Auger or radiative lifetime of an intermediate state $\Gamma^{-1}$ is short compared to a relevant phonon period $\tau_{ph}$, the lattice will not have time to respond to the creation of the core-hole, and the lattice will not contribute. Early photoemission studies of ionic insulators disproved this argument, measuring substantial spectral broadening linked to coupling between the core-hole and optical phonons \cite{Citrin1974a,Almbladh1977,Mahan1977}. The time-domain evolution of the core-excited state in these conditions, however, has yet to be explored.

%This argument, however, has proven to be largely false, as first evidenced by substantial broadening in the x-ray photoemission of ionic insulators \cite{Citrin1974a}. This broadening was linked to coupling between the core-hole and optical phonons \cite{Almbladh1977,Mahan1977}, demonstrating that nuclear motion plays a role for short- and long-lived core-holes alike \cite{Citrin1977}. A particularly intriguing situation was observed when $\Gamma$ is on the same order of magnitude as the phonon energy, which can occur in x-ray spectra: in this case, individual vibronic levels begin to overlap and their decays cannot any longer be considered independent. This effect was first observed in small molecules and named \textit{lifetime-vibrational interferences} \cite{Kaspar1979,GelMukhanov1977}, in the sense that vibronic levels open different interfering quantum paths contributing to the core-excitonic decay at a given energy. In solids, this phenomenon was also called \textit{incomplete phonon relaxation} \cite{Almbladh1977}. Although x-ray emission spectra have been understood in these conditions \cite{Gelmukhanov1999}, the real-time evolution of the intermediate state has yet to be observed.

In this Letter, attosecond XUV transient reflectivity spectroscopy is used to investigate the core-excited dynamics of magnesium oxide (MgO). In this ionic insulator, the low dielectric constant and the positive charge of Mg atoms mean that a Mg core-hole is very weakly screened. This contributes to the formation of core-level excitons - bound electron-hole pairs - at the Mg $\text{L}_{\text{2,3}}$ edge \cite{Lindner1986,OBrien1991}, i.e., close to transitions from Mg 2p states to the conduction band. First, four well-known core-level excitons are identified, and their linewidths are shown to be dominated by phonon broadening, indicating a substantial exciton-phonon coupling. Then, the core-level excitons are initiated by an XUV pulse and a short near-infrared (NIR) optical laser pulse probes their decay. The NIR/core-exciton interaction is interpreted using a few-level theoretical model, whose agreement with the experiment allows to identify two effects of the NIR pulse: (1) resonant and non-resonant coupling of the various excitonic states, and (2) creation of light-induced features that are interpreted as the manifestation of dark excitonic states. The decays of the first two lower-energy core-excitons are extracted, showing coherence lifetimes below 3 fs. Finally, the link between these features and exciton-phonon coupling is discussed.

The experiment was performed on a single-crystalline commercial sample of MgO (100) (MTI corporation) featuring industrial grade polishing (RMS $<$10 \si{\angstrom}) and a crystalline purity $>$99.95\%. Given that the samples are thick, 0.5 mm, the investigation mandates the use of Attosecond Transient Reflectivity Spectroscopy (ATRS) \cite{Kaplan2018} which does not require thin films, contrary to transmission experiments. Broadband pulsed XUV attosecond radiation covering a continuous span of 20-70 eV is produced by High Harmonic Generation (HHG) in argon using 480 $\mu$J, sub-5 fs NIR pulses centered at 750 nm at a repetition rate of 1 kHz. After filtering the NIR light with a 100 nm thick aluminum film, the XUV pulse is reflected by the sample at an incidence angle of 66\degree{ }from normal. The reflected light is then dispersed and imaged onto a CCD camera (Princeton Instruments). The spectral resolution at 50 eV is $\sigma = 25$ meV as determined using atomic absorption lines. In the time-resolved experiment, a strong 4.5 fs pulse of NIR light perturbs the system following the core-hole creation by the attosecond pulse. At each time-delay between the XUV and NIR pulses, the reflectivity $R_\text{on}$ (resp. $R_\text{off}$) is measured with 100 ms integration time with the NIR beam on (resp. off). The transient reflectivity, $\frac{dR}{R}=\frac{R_\text{on}-R_\text{off}}{R_\text{off}}$, is computed and averaged over 100 full time-delay scans. The slow drift of the pump-probe delay is stabilized over several hours using periodic reference measurements \cite{Jager2018}. The XUV and NIR pulses are s- and p-polarized with respect to the sample surface, respectively  \footnote{s-polarized XUV and p-polarized NIR (with respect to the sample) were used throughout this study, as they respectively provide more XUV reflectivity and less background noise}.

\begin{figure}[!h]
 \includegraphics[width=\linewidth]{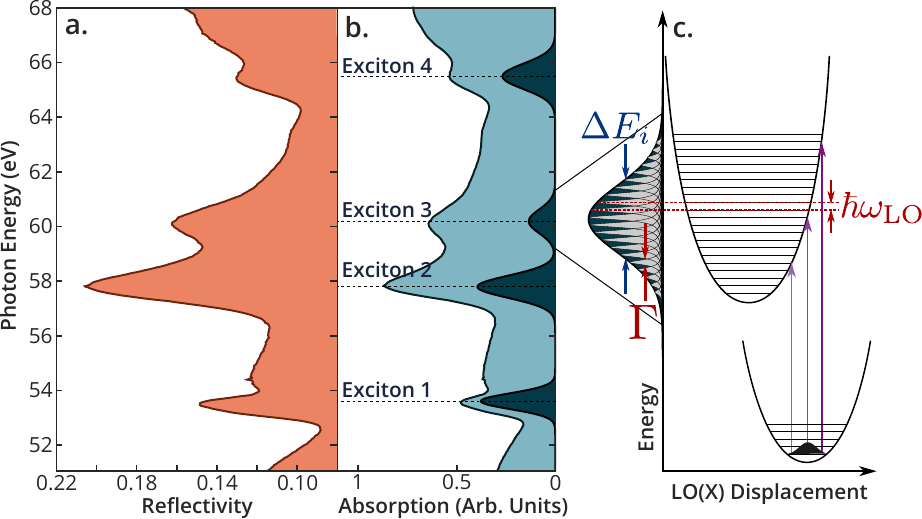}
 \caption{\label{fig1} Absolute reflectivity at the Mg $\text{L}_{\text{2,3}}$ edge in MgO(100). (a) Measurement at 66\degree{ }from normal incidence; (b) Absorption coefficient obtained by Kramers-Kronig analysis (light blue) , with 4 exciton peaks visible (dotted lines). Each exciton peak can be fit to a Gaussian distribution (dark blue). (c) Schematic of linear exciton-phonon coupling: the core-excited state potential energy surface is shifted from the ground-state potential, along the LO(X) phonon coordinate. In the strong coupling regime, the absorption lineshape of the $i$-th exciton is a Gaussian of width $\Delta E_i$. It is composed of a series of Lorentzians of width $\Gamma$, spaced by $\hbar \omega_0$, the phonon energy.}
\end{figure}

Fig. \ref{fig1}a shows the XUV reflectivity of the MgO crystal in the absence of the NIR pulse. During the measurement, the reflectivity of MgO and a calibrated gold mirror are taken sequentially allowing signal normalization and thus obtaining the absolute reflectivity of MgO. The reflectivity is a combination of both the real and imaginary parts of the refractive index, which makes its direct interpretation difficult \cite{Kaplan2019}. We therefore extract the absorption spectrum, shown in Fig. \ref{fig1}b, via a Kramers-Kronig (KK) analysis: the data are padded with literature data \cite{ROESSLER2007} covering the infrared to X-ray regions, which allows performing the KK integral over a wide frequency range. Four exciton peaks are measured, on top of a continuum resulting from core-to-conduction band transitions. The exciton energies match with previous absorption and electron energy-loss measurements within 0.5\% or less \cite{Henrich1976}. After removal of the continuous background, the absorption profiles can be fitted to Gaussian lineshapes. These are shown on the right side of Fig. \ref{fig1}b., and the Gaussians have full-widths at half maxima (FWHM) of $\Delta E_{1-4}$ = 0.68, 0.95, 1.03 and 1.10 eV, from low to high energy, respectively. The Mg 2p hole involved here is known to be filled almost solely by intra-atomic Auger processes \cite{Citrin1974}. This electronic decay process alone would thus give a narrow Lorentzian lineshape with $\Gamma_{\text{2p}}$ = 30 meV \cite{Citrin1977a}\footnote{For the purposes here we follow Citrin and Hamann \cite{Citrin1977} and assume that the Mg 2p lifetime in MgO is comparable to the lifetime in Mg metal.}, at clear variance with the experiment. In addition, since the sample is monocrystalline, we can safely disregard inhomogeneous broadening that would be caused by site-to-site fluctuations of the exciton energy in a disordered sample. Therefore, the broad Gaussian linewidths must be caused by strong exciton-phonon coupling, as represented on Fig. \ref{fig1}c and as already recognized for the lowest energy exciton \cite{OBrien1991}. Mahan \cite{Mahan1980} showed that in crystals with halite structure, such as MgO, core-excitons couple mainly to the longitudinal optical (LO) phonon at the X point, which is therefore the coordinate of Fig. \ref{fig1}c. For MgO, this phonon has an energy of $\omega_{LO}$=60 meV \cite{Agarwal1981}. The absorption lineshape is thus determined by the Franck-Condon factor of the vibronic excitation and the lifetime broadening of the core-excitons through Auger decay. After deconvolution of the Auger linewidth and the spectral resolution of the experiment, the phonon broadening of each exciton is equivalent to 11.1, 15.7, 16.7 and 17.9 phonons, from low to high energy, respectively. This strong coupling is the result of the ionicity of MgO: the principal way of screening the hole on the Mg cation is the motion of neighboring anions.

\begin{figure}[ht]
 \includegraphics[width=\linewidth]{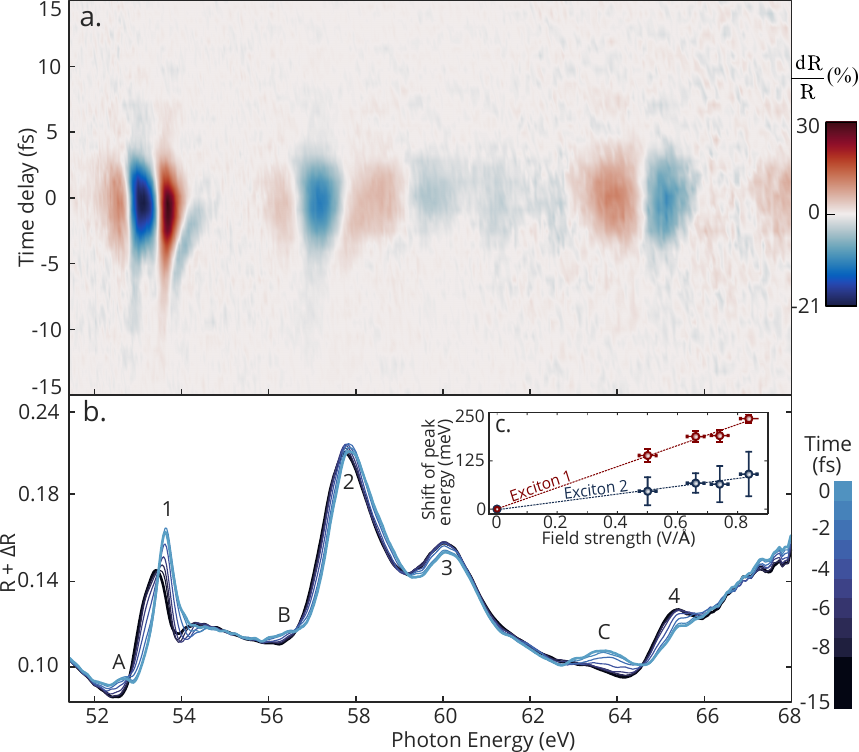}
 \caption{\label{fig2}Attosecond-resolved dynamics of excitons in MgO. (a) Time-delay dependent transient reflectivity, measured with a probe field strength of $0.84 \pm 0.01$ \si{V/\angstrom}, or \SI{9.3 \pm 0.2 e12}{W/cm^2}. (b) Absolute reflectivity, obtained by adding the static and differential ones, for pump-probe delays from -15 to 0~fs. Features corresponding to excitonic shifts (1 to 4) and light-induced features (A to C) are indicated. (c) Energy shift of excitons 1 (red) and 2 (blue) at $t = 0$ as a function of the NIR field strength. The dotted lines are linear fits, giving shifts of 274 and 101 meV/(\si{V/\angstrom}), respectively.}
\end{figure}

The transient reflectivity of the system is shown in Fig. \ref{fig2}a, where negative times mean the XUV comes first. Hence, the changes at negative delays around the four exciton energies are a direct time-domain observation of the core-excited state decay. Remarkably, the transient reflectivity signal lasts only for a few femtoseconds, indicating that the excitonic state experiences a very fast dephasing mechanism. This observation demonstrates the ability of attosecond spectroscopy to capture the dynamics of extremely short-lived core-excited states in the condensed phase, as was previously done for longer-lived states in atomic species \cite{BeckNJP2014,Cao2016,Beaulieu2017a,Ott2013a,Ott2014}. We note that the absence of longer-lived transient features, commonly observed in semi-conductors \cite{Zurch2017a,Schlaepfer2018,Kaplan2018}, allows us to disregard carrier excitation across the 7.8 eV bandgap of MgO.
%Our experiment did not reveal any change in the dynamics when the polarization angle of either the pump or probe beam was varied, indicating predominantly isotropic exciton oscillator strengths.  Even at the strongest pumping power, no signal due to accumulation of heat was observed.
%Instead, the observed changes are interpreted as the result of NIR-induced couplings of the states of the system during the core-hole relaxation. This translates into a modification of the XUV-triggered free-induction decay, a well-studied effect in atomic attosecond studies \cite{BeckNJP2014,Cao2016,Beaulieu2017a}.    %Therefore, s-polarized XUV and p-polarized VIS/IR (with respect to the sample) were used throughout this study, as they respectively provide more XUV reflectivity and less background noise.

Figure \ref{fig2}b shows the time-dependent absolute reflectivity of MgO. We identify three types of features with distinct behaviors: 1) the two lowest excitons, marked as 1 and 2, experience a blue-shift near zero-delay; 2) the reflectivities of the two higher energy excitons, 3 and 4, decrease; 3) new features, marked as A to C, emerge at energies where no signal was initially present. A KK analysis of this data shows that the first two trends are consistent with the attosecond transient absorption experiment of Moulet et al. \cite{Moulet2017} in Si$\text{O}_2$, confirming the excitonic nature of peaks 1 to 4. On the other hand, features A, B and C show a strikingly different behavior in that they do not seem to originate from either of the core-excitons, but rather appear only in the presence of the strong NIR field.

The experiment was repeated while varying the strength of the NIR electric field, resulting in a linear increase of the blueshifts of excitons 1 and 2 (Fig. \ref{fig2}c), with significantly different polarizabilities between excitons 1 and 2. In addition, this behavior is distinct from the pure optical Stark effect of a free exciton, which for large pump detunings would give a redshift increasing linearly with the NIR intensity \cite{Combescot1992}. NIR-driven couplings between excitonic levels must therefore play an important role here. Finally, our experiment did not reveal any change in the dynamics when the polarization angle of either the pump or probe beam was varied, indicating predominantly isotropic exciton oscillator strengths.  Even at the strongest pumping power, no signal due to accumulation of heat was observed.

We attempt to disentangle the two types of exciton dynamics driven by the phonon- and the NIR-driven couplings, respectively, via a simple model for the core-exciton system interacting with a two-color XUV and NIR pulse. Given that the excitonic states originate in spatially localized bound electron-hole pairs, they are similar in character to atomic states. We thus start by constructing a four-level system consisting of the ground state, 0, excitons 1 and 2 (which exhibit the most intense behavior), and a so-called dark exciton state $d$ that has no dipole-coupling to the ground state but can couple by the optical NIR pulse to both exciton states. Although we do not predetermine the energy of the dark state (we will later extract this energy by fitting the model calculations to the experimental results), we do assume the dark state is in the vicinity of the other exciton states. Solving the time-dependent Schr\"odinger equation (TDSE) for this system yields a time-dependent dipole moment of the form: 
%
%In order to extract information about the core-excitonic decay from these findings, we undertake a theoretical modeling of the ATRS spectrogram. First, we note that excitons hold a close resemblance to atomic systems: much like an hydrogen atom, each of them is constituted of a bound electron-hole pair with the electron being more delocalized. Consequently, each exciton is described as a discrete level embedded in a continuum of states. In what follows, we focus on excitons 1-2 and features A-B, which show the most distinct behavior. Looking at the orbital character of the conduction band \cite{Elsasser2001}, and considering only dipole-allowed transitions from the Mg 2p core-level, exciton 1 can be assigned s-like character, while exciton 2 has mixed s/d character. Therefore, the NIR can couple them with a p-like dark exciton whose existence is first postulated, and will be verified later. 
%
\begin{equation}
\begin{split}
D(t,\tau) &\propto  2 \Re\big(c^{*}_0c_1\mu_{0,1} e^{i\phi_1(t,\tau)} + c^{*}_0c_2 \mu_{0,2} e^{i\phi_2(t,\tau)} \\
&+ c^{*}_1c_d \mu_{1,d} e^{i\phi_1(t,\tau)}  + c^{*}_2c_d
\mu_{d,2} e^{i\phi_2(t,\tau)}\big),
\end{split}
\end{equation}
where the subscripts 0, 1, 2, $d$ denote  the four states, $\tau$ is the XUV-NIR delay, $\mu_{i,j}$ are the transition dipole moments, and $c_{i}$ are the delay- and time-dependent amplitudes of the four states. 

The exciton phases $\phi_1$ and $\phi_2$ are added to the dipole in order to account for three dynamical effects that go beyond the 4-level TDSE model, at the phenomenological level: $\phi_i(t,\tau) = i\Gamma_{2\mathrm{p}} t + \phi_L(t,\tau) + \phi_{\mathrm{ph},i}(t)$, where $\Gamma_{2\mathrm{p}}=30$ meV \cite{Citrin1977a} is the 2p core-hole Auger decay rate, $\phi_L(t)$ is the AC Stark phase imposed by laser-dressing of the loosely bound exciton states, and $\phi_{\mathrm{ph},i}(t)$ is due to the phonon coupling. The AC Stark phase is proportional to the energy shift $U_p$ experienced by a free electron in an oscillating field:  
\begin{equation}
\phi_L(t,\tau) = -\beta \int_0^t U_p(t',\tau)dt'
%\int_0^t{E_{\mathrm{NIR}}^2(t',\tau)\mathop{}\mathrm{d}t'},
\end{equation}
and $\beta$ is determined via fitting as described below. The exciton-phonon coupling phase is given by \cite{Mahan1977}:
\begin{equation}
\begin{split}
\phi_{\mathrm{ph},i}(t) 
    =& i\frac{M_i^2}{\omega_{LO}^2}\left[(2N+1)(1-\cos\omega_{LO} t) \right.\\
    &\left. -i(\omega_{LO} t-\sin\omega_{LO} t) \right],
\end{split}
\label{eq:phonon}
\end{equation}
where $N$ is the thermal phonon population, $\omega_{LO}=60$~meV is the LO phonon energy \cite{Agarwal1981}, and $M_i$ is the exciton-phonon coupling constant, to be determined by the fit. %This expression was first derived by Mahan in the context of understanding photoemission from metallic lithium. 
 Finally, the absorption spectrogram is calculated as $A(\omega,\tau) = - \omega \Im[E^*_{\mathrm{XUV}}(\omega) \tilde{d}(\omega,\tau)]$, with $E_{\mathrm{XUV}}(\omega)$ and $\tilde{D}(\omega,\tau)$ the Fourier transforms of the XUV electric field and $D(t,\tau)$, respectively \cite{Gaarde2011}. This is compared to the experimental delay-dependent absorption spectrum, extracted from a KK analysis of the measured transient reflectivity and displayed in Fig. 3(a-b). 
 
\begin{figure}[!h]
 \includegraphics[width=.86\linewidth]{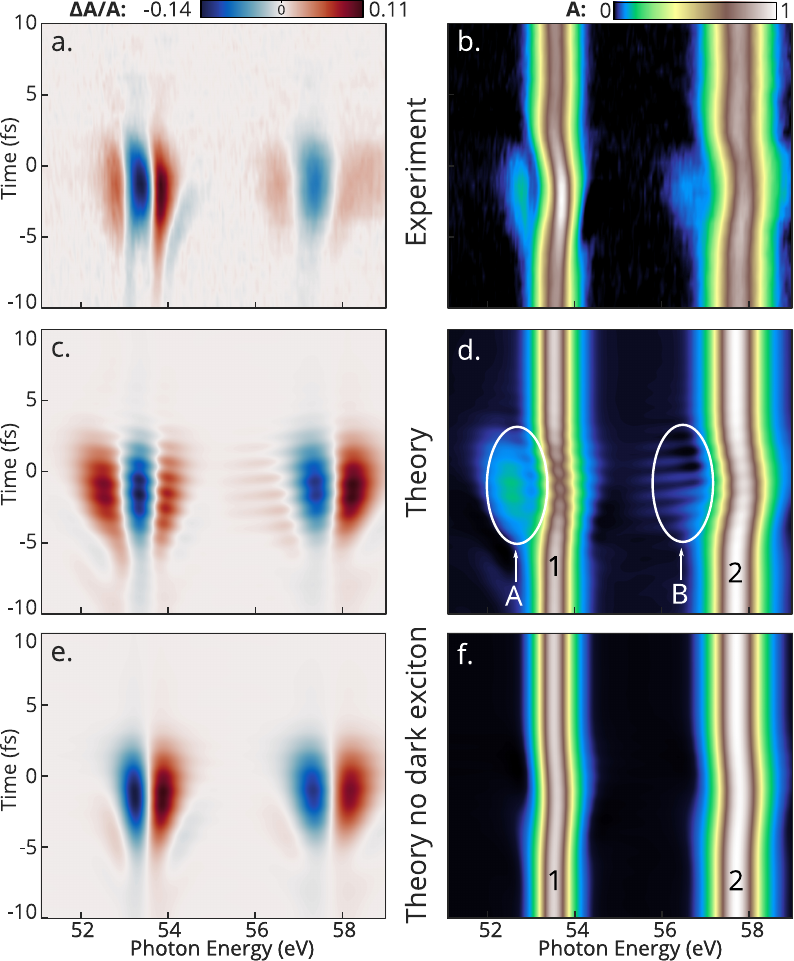}
 \caption{\label{fig3} Transient experimental (a-b) and calculated (c-f) absorption spectra. The experimental spectrograms are obtained from KK analysis of Fig. \ref{fig2}a. (a,c,e) and (b,d,f) show changes in absorption and absolute absorption, respectively. Results from calculations are normalized to the same scale as experiments. (e,f) show a calculation in which the dark state is omitted.}
\end{figure}

 The parameters used in the model are extracted in two stages. 
 We first fit the calculated XUV-alone spectrum to the experimental spectrum  to obtain the phonon coupling constants $M_1 = 0.2663 \pm 0.0080 \;\text{eV}$ and $M_2 = 0.3667 \pm 0.0071 \;\text{eV}$, and $\mu_{0,2} = -0.0818 \pm 0.0013$ a.u. (when using $\mu_{0,1} = 0.05$ a.u. as an overall scaling parameter). The parameters characterizing the laser-driven couplings are found by fitting the measured spectrum in the interval [51.1 eV, 59.1 eV], at overlap ($\tau =0$), to that calculated by using the measured NIR pulse shape as input, with a peak intensity of $2\times 10^{12}$~W/cm$^2$, and averaging over approximately one NIR optical cycle around $\tau =0$. A full delay-dependent calculation is then performed using the extracted parameters. The main discrepancy in the result is the appearance of sub-cycle oscillations in the theoretical absorption, which are not resolved in the experimental delay scan. We speculate that this could be due to experimental factors such as time-delay jitter, or phase slip effects between NIR and XUV, possibly important in our reflectivity geometry. For this reason, a half-cycle moving average is added to the calculation. Even though the oscillations do not completely disappear, this allows a clearer comparison, as shown in Fig. \ref{fig3}(c-d).

The comparison between experiment and theory is otherwise good, with the dynamics of both the excitonic blueshifts and the A-B features reproduced well. Although the fitted parameters can only be extracted with large uncertainties, there are several important lessons learned from the comparison: First, the presence of the dark state is crucial for the appearance of features A and B. This is demonstrated in Fig. 3(e-f) which shows the calculated spectrogram in the absence of the dark state. The energy of the dark state is found to be $54.4\pm 0.3$~eV; {\it i.e.} located between excitons 1 and 2, so that features A and B are one NIR photon energy below and above the dark state, respectively. This suggests that A and B are light-induced states (LISs) that can be understood as the intermediate state in two photon XUV+NIR transitions from the ground to the dark state. LISs are well-known from atomic transient absorption studies \cite{Chen2012,JustineBell2013,Chini2013} but have not previously been observed in the solid state. We find that the dark-state coupling matrix elements are large (similar to those between atomic states \cite{Wu2016}), and comparable in magnitude to each other, $\mu_{1,d} = 3.8 \pm 0.5$~a.u. and $\mu_{d,2} = -3.7 \pm 1.0$~a.u, \footnote{In the fit we cannot distinguish between whether the signs of $\mu_{1,d}$ and $\mu_{d,2}$ are the same or opposite. Choosing initial conditions of same signs yields a dark state energy and matrix elements of the same order of magnitude.} and that $\beta = 1 \pm 0.5$.

\begin{figure}[!h]
 \includegraphics[width=.9\linewidth]{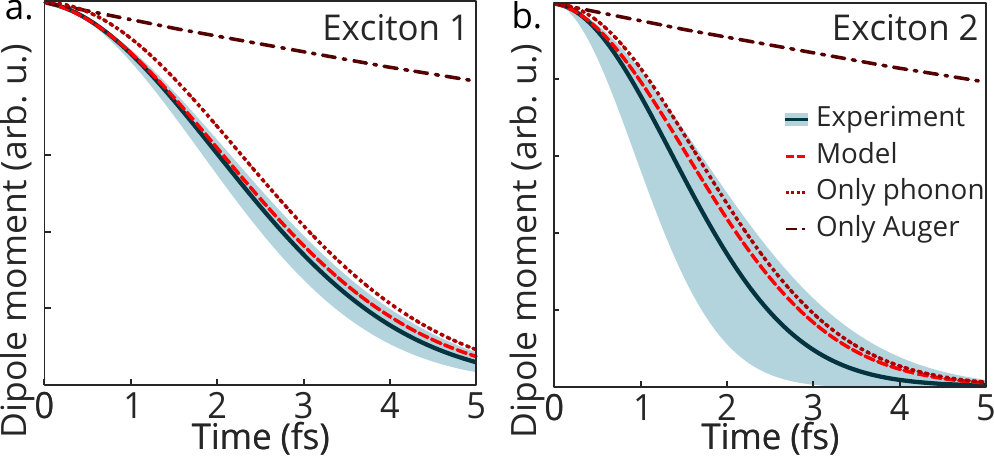}
 \caption{\label{fig4} Dipole decays obtained from fitting the delay-dependent energy position of exciton 1 (a) and 2 (b) (full blue lines). The shaded areas represent the 95\% confidence interval of the fitting in addition to a $\pm 0.5$ fs uncertainty on the NIR pulse duration. The experimentally obtained dipole decay is compared with the full model (dashed red line), the model with only the phonon contribution (dotted line) or with only the Auger decay (dash-dotted line).}
\end{figure}

The fact that the NIR-induced attosecond excitonic dynamics are well described using concepts from atomic physics highlights the striking similarity between excitons and isolated atoms. However, the way these quasiparticles decay is remarkably different for the solid versus atoms. Indeed, the Mahan model that we use to describe exciton-phonon coupling (phonon phase in Eq.~\ref{eq:phonon}) predicts a Gaussian decay of the dipole moment for short times: $e^{-\Im(\phi_{\text{ph}})} \propto e^{-M_i^2 t^2}$ for $\omega_{LO} t \ll 1$. Thus, in the framework of this model, the excitonic coherence lifetime is directly reduced by the strength of the exciton-phonon coupling. To validate the use of this model in our case, we compare its result to the experimental data. The center energy of excitons 1 and 2 is measured as a function of pump-delay, and the result is fitted as the convolution of the NIR intensity and an unknown dipole decay for each exciton. Fig. \ref{fig4} shows that the comparison of this fit to the theoretical model is reasonable, yielding coherence decay times at half maximum of $2.3\pm 0.2$ fs and $1.6\pm 0.5$ fs. Including either just the Auger or just the phonon dephasing decay (Fig. \ref{fig4}) shows that the exciton decay is dominated by the exciton-phonon coupling.

Returning now to the static absorption profiles of Fig. \ref{fig1}, the Gaussian linewidths $\Delta E_1$ and $\Delta E_2$ correspond to dephasing times of $2.6\pm 0.3$ and $1.9\pm 0.4$ fs, respectively. The similarity between these values and the dephasing times measured in the time domain indicates that exciton lifetimes in MgO are mainly governed by $\Delta E_i$, the bandwidth of the vibronic wavepacket produced during the absorption process. We interpret this finding as an analog of \textit{vibrational-lifetime interferences} observed in small molecules \cite{Kaspar1979,GelMukhanov1977}: since the lifetime width (in our case, the Auger width) is on the same 
order of magnitude as the phonon frequency, vibronic levels overlap and open interfering quantum paths contributing to the core-excitonic decay at a given energy. Thus, the effect of phonons on the core-excited state decays cannot be gauged by the phonon period (69 fs, in our case), but rather by (i) the strength of the exciton-phonon coupling and (ii) the ratio between lifetime width and phonon frequency. The requirements for dominant phonon dephasing are thus relatively simple to meet; in turn, we expect the findings presented here to have important ramifications in the time-domain studies of many similar short-lived core-excited states.  %This is verified for MgO, in which the highly ionic Mg-O bond forces the anion to move to screen the core-hole, and for which $\Gamma_\text{2p}/\omega_\text{LO} \approx 2$.
%This can be seen as a manifestation of  in the time domain, wherein the excitonic decay at each energy can occur through a number of interfering quantum pathways, leading to a rapid decay. $\Delta E_i$ results from the shape of the potential energy surfaces and the strength of the exciton-phonon coupling, i.e., the local displacement induced by the presence of the core-hole. Thus, it is seen that the effect of phonons on core-excited state decays cannot be gauged by the phonon period alone. Here, despite the LO(X) phonon period being 69 fs, its strong coupling with the core-hole is responsible for a sub-3 fs dephasing of the exciton states. Generally, vibrational-lifetime interferences will have a sizable influence on the time-domain decay if (i) the potential energy surface of the core-ionized state is displaced from the corresponding ground state equilibrium, and (ii) the lifetime width (in our case, the Auger width) is on the same order of magnitude as the phonon frequency, causing vibronic levels to interfere. This is verified for MgO, in which the highly ionic Mg-O bond forces the anion to move to screen the core-hole, and for which $\Gamma_\text{2p}/\omega_\text{LO} \approx 2$. %It is also the case for $\text{SiO}_\text{2}$, known to have short-lived excitons \cite{Moulet2017} and verifies $\Gamma_\text{2p}/\omega_\text{LO} \approx 2$ too.

In summary, we investigated the decay of core-excitonic states in MgO(100) using ATRS. Using a short NIR field to perturb the relaxation, excitonic blueshifts and light-induced states were identified as distinctive features of the exciton-NIR interaction. The coherence lifetimes of the excitons were found to be principally given by the inverse width of the vibronic wavepacket created in the core-excited state. This study furthers the understanding of light-induced modification of short-lived excitonic states and extends concepts of attosecond metrology to the solid state. It presents a direct measurement of an extremely short natural lifetime obtained using attosecond transient spectroscopy. Furthermore, the observation of substantial phonon dephasing is of compelling importance for future attosecond studies, which are making progress in the understanding of condensed phase systems such as solids and solvated molecules.

The authors would like to thank Eric M. Gullikson for performing the calibration of the gold mirror and are grateful to L. Barreau and H. Marroux for their comments on the manuscript. Experimental investigations were supported by the Defense Advanced Research Projects Agency PULSE Program Grant W31P4Q-13-1-0017, the U.S. Air Force Office of Scientific Research No. FA9550-14-1-0154, the Army Research Office MURI grant No. WN911NF-14-1-0383, and the W.M. Keck Foundation No. 046300. Theoretical work at LSU was supported by US Department of Energy, Office of Science, Basic Energy Sciences, under Contract No. DE-SC0010431.

%\bibliography{C:/Users/Romain/Documents/Library/library}
%\bibliography{references}
\bibliography{MgO_Draft_v5_forArxiv}
\end{document}